\documentclass[runningheads]{llncs}

\usepackage{yfonts}
\usepackage{amsmath}
\usepackage{amsfonts}
\usepackage{amssymb}
\usepackage{booktabs}
\usepackage{graphicx}
\usepackage[colorlinks]{hyperref}
\usepackage[switch]{lineno}
\usepackage{subcaption}
\usepackage[numbers,sort&compress]{natbib} 

\usepackage[T1]{fontenc}

\usepackage{color}

\newcommand{\ballot}{b}
\newcommand{\ballots}{\mathcal{B}}
\newcommand{\votevals}{\mathcal{V}}
\newcommand{\cvrs}{\mathcal{C}}
\newcommand{\outcome}{\mathcal{O}}
\newcommand{\schoice}{\mathbb{S}}
\newcommand{\ballotvar}{\mathcal{X}}
\newcommand{\cvrmargin}{V}
\newcommand{\propmargin}{v}
\newcommand{\mismatch}{mismatch-based}
\newcommand{\Mismatch}{Mismatch-based}
\newcommand{\MisMatch}{Mismatch-Based}

\title{Doing More With Less: Mismatch-Based Risk-Limiting Audits\thanks{Presented
at the 10th Workshop on Advances in Secure Electronic Voting
(Voting'25), on 18 April 2025.
Published in: FC 2025 Workshops, LNCS 15754, pp.\ 241--255, 2026,
\url{https://doi.org/10.1007/978-3-032-00495-6_13}.
This work was supported by the Australian Research Council
(Discovery Project DP220101012, OPTIMA ITTC IC200100009) and the U.S.\ National
Science Foundation (SaTC~2228884).}}

\author{
Alexander Ek       \inst{1}  \orcidID{0000-0002-8744-4805}  \and
Michelle Blom      \inst{2}  \orcidID{0000-0002-0459-9917}  \and
Philip B. Stark    \inst{3}  \orcidID{0000-0002-3771-9604}  \and
Peter J. Stuckey   \inst{4}  \orcidID{0000-0003-2186-0459}  \and
Vanessa J. Teague  \inst{5}  \orcidID{0000-0003-2648-2565}  \and
Damjan Vukcevic    \inst{1}  \orcidID{0000-0001-7780-9586}
}

\authorrunning{Ek et al.}

\institute{
Department of Econometrics and Business Statistics, Monash University, Clayton,
Australia
\\
\email{damjan.vukcevic@monash.edu}
\and
Department of Computing and Information Systems, University of Melboure,
Parkville, Australia
\and
Department of Statistics, University of California, Berkeley, CA, USA
\and
Department of Data Science and AI, Monash University, Clayton, Australia
\and
Australian National University and Thinking Cybersecurity Pty Ltd.
}

\begin{document}

\maketitle

\begin{abstract}
One approach to risk-limiting audits (RLAs) compares randomly selected cast
vote records (CVRs) to votes read by human auditors from the corresponding
ballot cards.  Historically, such methods reduce audit sample sizes by
considering \emph{how} each sampled CVR differs from the corresponding true
vote, not merely \emph{whether} they differ.  Here we investigate the
latter approach, auditing by testing whether the total number of mismatches
in the full set of CVRs exceeds the minimum number of CVR errors required
for the reported outcome to be wrong (the ``CVR margin'').  This strategy
makes it possible to audit more social choice functions and simplifies RLAs
conceptually, which makes it easier to explain than some other RLA
approaches.  The cost is larger sample sizes.  ``\Mismatch{}  RLAs'' only
require a lower bound on the CVR margin, which for some social choice
functions is easier to calculate than the effect of particular errors.
When the population rate of mismatches is low and the lower bound on the
CVR margin is close to the true CVR margin, the increase in sample size is
small.  However, the increase may be very large when errors include errors
that, if corrected, would widen the CVR margin rather than narrow it;
errors affect the margin between candidates other than the reported winner
with the fewest votes and the reported loser with the most votes; or errors
that affect different margins.

\end{abstract}


\section{Introduction}

Growing international skepticism of election outcomes emphasizes the need for
``evidence-based elections,'' which provide affirmative evidence that the
reported winners really won \citep{starkWagner12,appelStark20}.  Risk-limiting
audits (RLAs) using a demonstrably trustworthy record of the votes are a key
component of evidence-based elections.  RLAs limit the probability (risk) that
an incorrect reported outcome will become final by manually reading the votes
from randomly sampled records until either the sample provides sufficiently
strong statistical evidence that the reported winners really won or there has
been a full manual tabulation of the trustworthy vote record to determine the
correct winner.  RLAs are now mandatory or authorized in about fifteen U.S.\
states.

RLAs can be conducted using different sampling designs and different data,
depending on local laws, logistical constraints, and  the capability of
election technology~\citep{stark23ONEAudit}.  A particularly efficient approach
to RLAs, \emph{card-level comparison}, involves comparing a human reading of
the votes on validly cast ballot cards selected at random to the voting
system's record of the votes on each selected card, the \emph{cast vote record}
(CVR) for that card.  Methods for card-level comparison RLAs have been
developed for many social choice functions (the rules for determining who won
and how voters can express their preferences), including
first-past-the-post \citep{stark08a,stark10d,stark20},
supermajority \citep{stark08a,stark20},
instant-runoff voting (IRV)~\citep{blom2019raire,ek2023awaire},
D'Hondt \citep{StarkTeague2014},
Hamiltonian elections~\citep{voting21},
and approval, Borda, STAR-Voting, and arbitrary scoring rules \citep{stark20}.

There is so far no efficient RLA method for some social choice functions, such
as single transferable vote (STV) with more than two seats~\citep{voting22}
and certain Condorcet elections~\citep{voting23}.\footnote{%
A full manual re-tabulation of the votes is an RLA for
any social choice function, but it is inefficient.}
For some of these, it is possible to compute a \emph{lower bound} on the
\emph{CVR margin}---the minimum number of CVRs that must differ from the votes
on their corresponding ballot cards for the reported election outcome to be
incorrect.  This lower bound enables a straightforward audit: check
(statistically) whether the number of mismatches between CVRs and their
corresponding ballots is below this lower bound.  If so, the reported election
outcome is correct.  Here we define a \emph{\mismatch{} RLA}, which involves
testing (the complement of) this assertion, at significance level $\alpha$.
This method is statistically rigorous and broadly applicable.  It produces a
card-level comparison RLA whenever a lower bound on the CVR margin can be
computed, including social choice functions for which computing the CVR margin
is challenging but computing a lower bound is tractable.  For example,
computing the CVR margin for IRV is NP-hard~\citep{Xia} (but usually feasible
in practice~\citep{ecai2016}); similarly, for STV there exist methods for
calculating lower bounds~\citep{informsstv,blom2025stvmargin} for elections
where no efficient RLA approach is known.

\Mismatch{} RLAs make worst-case assumptions about every CVR error.
For instance, in a plurality contest, they treat the following errors as
equivalent:
\begin{itemize}
\item an error that changes a vote for the reported runner-up into a vote
    for the reported winner (the most serious error possible)
\item an error that changes an invalid vote into a vote for the reported
    winner (a less serious error)
\item an error that turns a vote for the candidate reported to have the
    fewest votes into a vote for another losing candidate (an error that
    might not be consequential for the outcome)
\item an error that changes a vote for the reported winner into a vote for
    the reported runner-up (an error that adds evidence that the outcome is
    correct).
\end{itemize}
This pessimism generally results in larger sample sizes than an audit that
considers the effect of each error---possibly requiring a full hand count when
a relatively small sample would suffice for a more nuanced method.  Using a
lower bound on the CVR margin (rather than the exact margin) exacerbates this
conservatism, depending on how how much slack the bound has.

\Mismatch{} RLAs require a CVR for each ballot card, ``linked'' to the card it
purports to represent, in order to compute mismatches.
Not every voting system can provide CVRs linked to physical cards.\footnote{%
Moreover, methods such as ONEAudit \citep{stark23ONEAudit} that create CVRs and
associate them to cards do not ``play nice'' with this approach because the
CVRs they produce have offsetting errors but possibly a high mismatch rate.}
Thus \mismatch{} RLAs may not help unless a jurisdiction uses central-count
optical scan systems or is willing to rescan precinct-tabulated cards.
Ballot-polling or batch-comparison audits are simply not possible using a
mismatch-based approach.

Here we examine the impact of the pessimistic treatment of mismatches using
simulated audits of IRV and two-candidate plurality contests, for which there
are tailored RLA methods \citep{blom2019raire,ek2023awaire}.  We find scenarios
where the workload is about the same, and others where \mismatch{} RLAs require
much larger samples.  \Mismatch{} auditing is perhaps most useful for social
choice functions such as STV, for which no better RLA method exists.

Our contribution is to explain the details of \mismatch{} RLAs and identify
scenarios in which they are helpful.  We also quantify their performance, using
simulations to show when they perform well and when they require much larger
samples than more precise methods.  Although the idea of \mismatch{} RLAs is
not new, we are not aware of any work that fleshes out the details or examines
their performance.

The paper is organized as follows.
\autoref{sec:prelim} provides definitions and notation.
\autoref{sec:mismatch-rla} defines and constructs
\mismatch{} RLAs and contrasts them with card-level comparison audits.
\autoref{sec:results} explores the relative efficiency of \mismatch{} RLAs
for various social choice functions using simulation.
\autoref{sec:conc} discusses the implications and directions for
future work.


\section{Terminology, Definitions, and Assumptions}
\label{sec:prelim}

We consider one contest at a time, but generalizing to an arbitrary number of
contests is straightforward \citep{stark10d,glazerEtal21}.
A \emph{ballot card} is a piece of paper.
(A \emph{ballot} comprises one or more ballot cards.)
There are $N$ validly cast ballot cards, each containing the contest in
question.
A \emph{vote} is a value derived from the marks a voter makes on a ballot card.
Depending on the social choice function, law, and regulation, some ways of
marking (or failing to mark) ballots produce ``invalid votes'' or
``non-votes,'' which the social choice function ignores; we call all of these
\emph{null votes}.

We are able to draw ballot cards at random from the full set of $N$ cards.
Each card is uniquely identifiable in some way, e.g., by its physical location
or through an identifying mark printed on the card;
there is a canonical ordering so that it makes sense to talk about the $i$th
ballot card.
Let $[N] := \{1, \ldots, N\}$.

The vote on the $i$th card is $\ballot_i$.
The set of possible values of $b_i$ is $\votevals$,
which includes valid and null votes.
The $N$-tuple of votes on the $N$ cards is $\ballots = (b_1, b_2, \dots, b_N)$.
If we select the $i$th card for audit, we observe $b_i$.
For each $i \in [N]$, we also have a
cast vote record $c_i$, committed to before the audit starts.
The $N$-tuple of $N$ cast vote records is $\cvrs = (c_1, c_2, \dots, c_N)$.
A \mismatch{} audit involves comparing $b_i$ to $c_i$ for a sequence of random
draws from $[N]$.

Many of these assumptions can be relaxed at the cost of some bookkeeping:
methods by \citet{stark20,stark23a,stark23ONEAudit} can deal with situations
where:
\begin{itemize}
\item the exact number $N$ of cards is unknown but an upper bound is available
\item the contest is not on every ballot card
\item the election official cannot account for all $N$ cards
\item a particular card is selected for audit but cannot be found
\item the number of cast vote records is not the same as the number of cards
\item the voting system cannot report a CVR linked to some cards.
\end{itemize}
The set of possible election outcomes is $\outcome$.
(Typically an outcome is a candidate, a party, an allocation of seats, or the
``yes'' or ``no'' position on some measure.)
The social choice function for the contest is $\schoice(\cdot)$, a mapping from
a tuple of elements of $\votevals$ (a tuple of votes), to an element of
$\outcome$ (an outcome).
The social choice function is assumed to depend on the tuple of votes only
though the multiset of votes---not, for example, on the order in which votes
are cast or tabulated, nor on auxiliary randomness to break ties.
The \emph{correct outcome} of the contest is $\schoice(\ballots) \in \outcome$,
the outcome that results from applying the social choice function to
$\ballots$, the actual votes on the ballot cards.
The \emph{reported outcome} is $\schoice(\cvrs) \in \outcome$,
the output of applying the social choice function to the CVRs $\cvrs$.

A common definition of \emph{margin of victory} for plurality contests is the
difference between the number of votes reported to have been received by the
winner and by the runner-up, which we call the \emph{vote margin}.

A more general notion of margin, applicable to any social choice function if
CVRs are available, is the minimum number of CVRs with a vote that differs from
the vote on the corresponding card if the reported outcome is wrong.  This
minimum, the \emph{CVR margin}, is taken over all possible true votes and all
subsets of the CVRs:
\begin{equation}
\cvrmargin(\cvrs) :=
    \min_{\ballotvar = (x_1, \ldots, x_N) \in \votevals^N}
        \min_{\mathcal{I} \subset [N]} 
            \{ |\mathcal{I}| : \schoice(\cvrs) \ne \schoice(\ballotvar)
            \text{ and }
            (c_i = x_i, \forall i \in [N] \setminus \mathcal{I})\}.
\end{equation}
Equivalently, $\cvrmargin(\cvrs)$ is the smallest radius of any
Hamming-distance ball centered at $\cvrs$ whose image under $\schoice(\cdot)$
contains at least two elements of $\outcome$:
\begin{equation}
    \cvrmargin(\cvrs) := \min \{n: \# \schoice(\mathbb{B}(\cvrs, n)) > 1 \},
\end{equation}
where $\mathbb{B}(\cvrs, n)$ is the Hamming ball of radius $n$ centered at 
$\cvrs$:
\begin{equation}
    \mathbb{B}(\cvrs, n) := \{ \mathcal{D} = (d_1, \ldots, d_N) \in \votevals^N
    \; \text{s.t.} \; \# \{i: c_i \ne d_i \} \leqslant n \}.
\end{equation}
This abstract definition is natural for audits that check whether $c_i = b_i$
for randomly selected values of $i$.

This definition tacitly assumes there is a CVR for every card.  It can be
modified to account for differences in the number of cards and CVRs;
alternatively, the number of CVRs and (an upper bound on) the number of cards
can be forced to be equal using methods by \citet{stark20,stark23a} to create
CVRs (with null votes) or delete CVRs (without changing the reported outcome)
before calculating the CVR margin using this definition.

The related normalized measure is the \emph{CVR margin proportion}, the CVR
margin divided by the number of cards
\begin{equation}
    \propmargin = \propmargin(\cvrs) := \frac{\cvrmargin(\cvrs)}{N}.
\end{equation}
This is similar to \emph{diluted margin} defined by \citet{stark10d}, but
the numerator uses the CVR margin rather than the vote margin.

In a plurality election, the vote margin is twice the CVR margin because a
change to a CVR can change the vote margin by up to two votes: changing votes
for the reported winner in $(\text{vote margin})/2$ CVRs into votes for the
reported runner-up yields a tie or a win for the runner-up.  In an IRV
election, finding the CVR margin is more complicated.

\begin{example}
    \label{ex:irv}
    Consider an IRV election with candidates
    $\outcome = \{ \text{Ali}, \text{Bob}, \text{Cal}, \text{Dee} \}$.
    Suppose $N=60$ cards were cast.
    Of the 60 CVRs,
    20 are (Ali),
    15 are (Dee, Ali, Bob),
    9 are (Cal, Dee),
    6 are (Bob, Cal, Dee),
    6 are (Ali, Cal), and
    4 are (Bob, Cal).
    The tally proceeds as shown in \autoref{tab:IRV1}: first Cal is eliminated
    and 9~votes flow to Dee; next Bob is eliminated and 6~votes flow to Dee and
    4 CVRs are exhausted; finally Ali is eliminated and Dee wins.
    The CVR margin is $\cvrmargin = 1$:
    changing one CVR from (Bob, Cal, Dee) to (Cal, Dee)
    changes the winner to Ali, as illustrated in \autoref{tab:IRV2}.
    The CVR margin proportion is $\propmargin = 1/60$.

    The \emph{last-round CVR margin} (the minimum number of CVRs that need to
    be changed in order to alter the outcome in the final round of the count)
    is 2, since Dee beats Ali by 4 votes: the CVR margin of an IRV election can
    be smaller than the last-round CVR margin.
    \qed
\end{example}

\begin{table}
\centering
\caption{IRV election from \autoref{ex:irv}.}
\label{tab:IRV}
    \begin{subtable}{0.45\textwidth}
    \caption{Original election}
    \label{tab:IRV1}
    \begin{tabular}{l|r|r|r}
        \toprule
        Cand. & Round 1 & Round 2 & Round 3  \\
        \midrule
        Ali  & 26 & 26  & 26  \\
        Bob  & 10 & 10  & --- \\
        Cal  &  9 & --- & --- \\
        Dee  & 15 & 24  & 30  \\
        \bottomrule
    \end{tabular}
    \end{subtable}~~~
    \begin{subtable}{0.45\textwidth}
    \caption{Election with one vote changed}
    \label{tab:IRV2}
    \begin{tabular}{l|r|r|r}
        \toprule
        Cand. & Round 1 & Round 2 & Round 3  \\
        \midrule
        Ali  & 26 & 26  & 41  \\
        Bob  &  9 & --- & --- \\
        Cal  & 10 & 19  & 19  \\
        Dee  & 15 & 15  & --- \\
        \bottomrule
    \end{tabular}
    \end{subtable}
\end{table}


\section{Risk-Limiting Audits Using Mismatches}
\label{sec:mismatch-rla}

Suppose we have a set of CVRs $\cvrs$ and we can compute a number $\cvrmargin^-
\leqslant \cvrmargin(\cvrs)$, a lower bound for the CVR margin.  Define
$\propmargin^- := \cvrmargin^- / N \leqslant \propmargin$.  Let $M = M(\cvrs)
:= \#\{i \in [N] : b_i \ne c_i \}$ be the number of CVRs with votes that do not
match the votes on the corresponding card, and let $m = m(\cvrs) := M(\cvrs) /
N$ be the \emph{mismatch rate}.  We can perform an RLA by testing (at
significance level $\alpha$) whether $M \geqslant \cvrmargin^-$, or
equivalently, whether $m \geqslant \propmargin^-$.  Rejecting that hypothesis
amounts to strong evidence that $\schoice(\ballots) = \schoice(\cvrs)$, i.e.,
that the reported outcome is correct.

We now develop a test of whether $M \geqslant V^-$ using the SHANGRLA
\citep{stark20} framework, which characterizes the correctness of outcomes in
terms of the means of bounded, nonnegative functions on votes called
\emph{assorters}.  An assorter uses the votes on a card and other information
(for instance, a list of CVRs) to assign a bounded, nonnegative number to that
card.

For a broad variety of social choice functions, the reported winners really won
if the mean of each of the lists that result from applying each assorter in a
collection of assorters to the true votes $\ballots$ is greater than 1/2.
(For IRV, there are many collections that suffice; the outcome is correct if
the means of all assorters in any of those collections are all greater than
1/2~\citep{ek2023awaire}.)

Testing whether the mean of an assorter is greater than $1/2$ can be done in
various ways~\citep{stark20,stark2023alpha,stark23ONEAudit}; the most relevant
for the current work is a card-level comparison audit, which is the most
efficient method when accurate comparison values for each card are available.
We first briefly review comparison audits in general.

Given an assorter $A$ with upper bound $u$, we can compare the assorter values
for the votes to a corresponding set of known
``reference values'' $x = (x_1, \ldots, x_N)$.
If CVRs $\cvrs = (c_1, \ldots, c_N)$ are available, the reference values might
be, for instance, $(x_1, \ldots, x_N) = (A(c_1), \ldots, A(c_N))$, but other
choices are possible~\citep{stark23ONEAudit}.

Let $\bar{A} := N^{-1} \sum_i A(b_i)$ and $\bar{x} := N^{-1}\sum_i x_i$.
The \emph{assorter margin} of $x$ is $\nu = \nu(x) := 2\bar{x} - 1$, twice the
amount by which the mean of the values of $x$ exceeds $1/2$.
Then
\begin{equation}
    \bar{x} - \bar{A} < \bar{x} - 1/2 = \nu/2
\end{equation}
iff $\bar{A} > 1/2$.
Comparison audits work by testing whether
$\overline{(x-A)} := N^{-1} \sum_i (x_i-A(b_i)) \leqslant \nu/2$,
as follows:
Define the \emph{overstatement assorter}
\begin{equation}
\label{eq:overstatement_assorter_def}
    B(b_i) := \frac{u + A(b_i) - x_i}{2u-\nu} \in [0, 2u/(2u - \nu)],
\end{equation}
and $\bar{B} := N^{-1} \sum_i B(b_i)$.
Then $\bar{A} > 1/2$ iff $\bar{B} > 1/2$.

If $A(b_i) = x_i$, the reference value for card $i$ matches the true value of
the assorter.
If $A(b_i) \ne x_i$, there is a \emph{mismatch} or \emph{discrepancy}.
If $A(b_i) > x_i$, the discrepancy is an \emph{understatement}: correcting it
widens the margin, so discovering the error adds to the evidence that the
reported outcome is correct.
If $A(b_i) < x_i$, the discrepancy is an \emph{overstatement}: correcting it
narrows the margin, so it reduces the evidence that the reported outcome is
correct.

For card-level comparison audits, $x_i = A(c_i)$, so the possible values of
$x_i$ are the same as the possible values of $A(b_i)$.  For plurality
(including multi-winner plurality), approval, and some other social choice
functions (including assorters used to audit IRV), the possible values of
$A(b_i)$ are 0, 1/2, and 1, so $u = 1$.  Thus for card-level comparison audits
of such social choice functions, the possible values of the numerator of
$B(b_i)$, $u + A(b_i) - x_i$ (i.e., $u + A(b_i) - A(c_i)$), are $0$, $1/2$,
$1$, $3/2$, and $2$.

In a two-candidate plurality contest, Ali vs.\ Bob, where Ali is the reported
winner, correctness of the outcome is determined by a single assorter $A$:
\begin{equation}
    A(b_i) = \left \{
    \begin{array}{ll}
        1,   & b_i = \text{Ali}       \\
        1/2, & b_i = \text{null vote} \\
        0,   & b_i = \text{Bob}.
    \end{array}
    \right .
\end{equation}
Ali got more votes than Bob iff $\bar{A} > 1/2$.
For this assorter, $u = 1$, and the numerator of
\autoref{eq:overstatement_assorter_def} takes the following values:
\begin{equation}
u + A(b_i) - x_i =
    \left \{
    \begin{array}{lll}
      0, & b_i = \text{Bob}, c_i = \text{Ali}       & \text{(2-vote overstatement)} \\
    1/2, & b_i = \text{Bob}, c_i = \text{null vote} & \text{(1-vote overstatement)} \\
    1/2, & b_i = \text{null vote}, c_i = \text{Ali} & \text{(1-vote overstatement)} \\
      1, & b_i = c_i                                & \text{(match)} \\
    3/2, & b_i = \text{null vote}, c_i = \text{Bob} & \text{(1-vote understatement)} \\
    3/2, & b_i = \text{Ali}, c_i = \text{null vote} & \text{(1-vote understatement)} \\
      2, & b_i = \text{Ali}, c_i = \text{Bob}       & \text{(2-vote understatement)}.
    \end{array}
    \right .
\end{equation}

We now derive an assorter for \mismatch{} RLAs.
Let $v'$ be any nonnegative number less than or equal to $\propmargin$.
Define the \emph{mismatch assorter}:
\begin{equation}
   C(b_i) := \frac{1 - 1_{b_i \ne c_i}}{2-2v'} \in \{0, 1/(2 - 2v')\}.
\end{equation}
If $\bar{C} := N^{-1} \sum_i C(b_i) > 1/2$,
then $m < \propmargin' \leqslant \propmargin$,
$M(\cvrs) < \cvrmargin$, and the reported outcome is correct.
Note that this assorter requires that CVRs are ``linked'' to cards, in order to
determine whether $b_i \ne c_i$.

The mismatch assorter is a special case of a supermajority assorter
\citep[\S2.3]{stark20}:
the outcome is correct if the fraction of matching CVRs is greater than 100\%
minus the CVR margin proportion.
(It is simpler than a supermajority assorter in that it does not take the
value~$1/2$.)
The mismatch assorter is structurally similar to an overstatement assorter,
with $u = 1$ and $v' = \nu/2$
(recall that for plurality contests, the vote margin is twice the CVR margin;
SHANGRLA in effect reduces elections to a collection of two-candidate plurality
contests).
However, the largest value $C(b_i)$ can attain is
$1 / (2 - 2v')$, while $B(b_i) \leqslant 2u / (2u - \nu)$;
this has implications discussed below.


\section{Results}
\label{sec:results}

\subsection{Algorithms and Parameter Settings}

Testing the hypothesis $\bar{C} \leqslant 1/2$ yields a \mismatch{} RLA with
risk limit no greater than the significance level of the test.  There are many
ways to perform that test.  Here we use a sequential random sample of values of
$C(b_i)$ without replacement and the ALPHA test supermartingale
\citep{stark2023alpha}.

We investigated the performance of \mismatch{} audits using simulations based
on the current implementation of SHANGRLA\footnote{%
\url{https://github.com/pbstark/SHANGRLA}, visited 23~January 2025.}
with minor modifications.
The ALPHA test in SHANGRLA is parametrized by a function $\eta_j$ that can be
thought of as an estimator of the mean of the values of the assorter that
remain just before the $j$th random draw.
Using the true mean minimizes the expected sample size when the assorter has
only two possible values, as is the case for $C_i$; it yields
Wald's sequential probability ratio test~\citep{stark2023alpha}.
Of course, the true mean is unknown: the default function $\eta_j$ in the
current implementation is a truncated shrinkage estimator with several tuning
parameters, including an initial guess $\eta_0$, a weight $d$ for the initial
guess relative to the sample, and a ``guardrail'' parameter $c$ that controls
how quickly $\eta_j$ can approach $1/2$ as the sample grows.

For the mismatch assorter we used $d = 100$ (the default), and $\eta_0 =
0.999/(2 - 2\propmargin)$, corresponding to an assumed rate of $10^{-3}$
erroneous CVRs.  Because the truncated shrinkage estimator produced values of
$\eta_j$ that were larger than optimal for \mismatch{} audits, we added a
constraint to keep $\eta_j$ from approaching its upper bound too quickly (the
mirror image of the ``guardrail'' that keeps $\eta_j$ above $1/2$) and used a
smaller value of $c$.
For card-level comparison audits (overstatement assorters), we used the COBRA
estimator $\eta_j$ \citep{COBRA}, which is parametrized by an assumed rate of
2-vote overstatements; we used $10^{-5}$.  Differences between the functions
used for $\eta_j$ in the two types of audit may account for some of the
observed performance differences.
Code is available at: \url{https://github.com/aekh/margin-audit}.

\subsection{Simulations}

We ran several experiments to quantify the performance of \mismatch{} RLAs and
compared their performance to that of existing RLA methods.  For each
experimental condition, we created $N$-tuples of $\cvrs$ of CVRs and $\ballots$
of votes, then conducted 1,000 independent audits using those pairs and
summarized the performance of each method by the mean of the 1,000 sample
sizes.

\begin{table}[!t]
\centering
\caption{\textbf{Performance of \mismatch{} RLAs.}
The mean sample size (columns~3--8) for a given CVR margin proportion
($\propmargin$), number of CVRs ($N$) and mismatch rate ($m$).
All entries show the mean sample size in 1,000 simulated audits.
`F' means every simulation required a full manual tabulation.}
\label{tab:mismatch-audit-results}
\begin{tabular}{lrrrrrrr}
\toprule
& &
\multicolumn{6}{c}{Mismatch rate ($m$)} \\
\cmidrule(l){3-8}
$\propmargin$ & $N$ & None & 0.0001 & 0.0003 & 0.001 & 0.003 & 0.01 \\
\midrule
0.001 & 10k & \phantom{0}
          2,835 & 3,587 &  5,332 &  9,751 & F & F \\
 &  50k & 3,236 & 4,681 &  9,208 & 47,850 & F & F \\
 & 100k & 3,292 & 4,958 & 10,505 & 95,809 & F & F \\
\midrule
0.002
 &  10k & 1,509 & 1,757 & 2,382 &  5,629 &  9,947 & F \\
 &  50k & 1,612 & 1,958 & 2,894 & 10,649 & 49,649 & F \\
 & 100k & 1,626 & 2,010 & 3,030 & 12,654 & 99,410 & F \\
\midrule
0.003
 &  10k & 1,022 & 1,141 & 1,428 & 2,962 &  9,640 & F \\
 &  50k & 1,068 & 1,225 & 1,605 & 3,991 & 47,621 & F \\
 & 100k & 1,074 & 1,237 & 1,634 & 4,274 & 95,912 & F \\
\midrule
0.006
 &  10k & 515 & 550 & 626 &   985 & 3,106 &  9,962 \\
 &  50k & 526 & 562 & 652 & 1,069 & 4,369 & 49,678 \\
 & 100k & 527 & 572 & 658 & 1,068 & 4,733 & 99,552 \\
\midrule
0.01
 &  10k & 308 & 323 & 345 & 469 & 1,022 &  9,557 \\
 &  50k & 312 & 327 & 359 & 484 & 1,109 & 47,144 \\
 & 100k & 312 & 329 & 360 & 483 & 1,143 & 95,137 \\
\midrule
0.02
 &  10k & 152 & 156 & 162 & 190 & 294 & 1,321 \\
 &  50k & 153 & 157 & 163 & 194 & 296 & 1,570 \\
 & 100k & 153 & 158 & 164 & 193 & 295 & 1,583 \\
\midrule
0.03
 &  10k & 101 & 103 & 105 & 117 & 160 & 443 \\
 &  50k & 101 & 103 & 106 & 118 & 152 & 465 \\
 & 100k & 101 & 103 & 105 & 118 & 156 & 463 \\
\midrule
0.06
 & (all) & 50 & 50 & 51 & 54 & 61 & 106 \\
\midrule
0.1
 & (all) & 30 & 30 & 30 & 31 & 34 &  45 \\
\bottomrule
\end{tabular}
\end{table}

\subsection{Performance of \MisMatch{} RLAs}
\label{sec:mismatch-audit-performance}

The expected sample size of a \mismatch{} RLA depends on:
(i)~the number of CVRs $N$,
(ii)~the CVR margin proportion~$v$ (or lower bound~$v^-$), and
(iii)~the mismatch rate $m$.
\autoref{tab:mismatch-audit-results} shows the results for all combinations of
the following:
$N \in \{10^4, 5\times 10^4, 10^5\}$;
$\propmargin \in \{ 0.001, 0.002, 0.003, 0.006, 0.01, 0.02, 0.03, 0.06, 0.1\}$;
$m \in \{0, 0.0001, 0.0003, 0.001, 0.003, 0.01\}$.
Unsurprisingly, when $m \gtrapprox \propmargin$,
a \mismatch{} RLA usually requires a full manual tabulation.

\subsection{Two-candidate Plurality Contests}

\begin{figure}[!t]
\centering
\includegraphics[width=\linewidth]{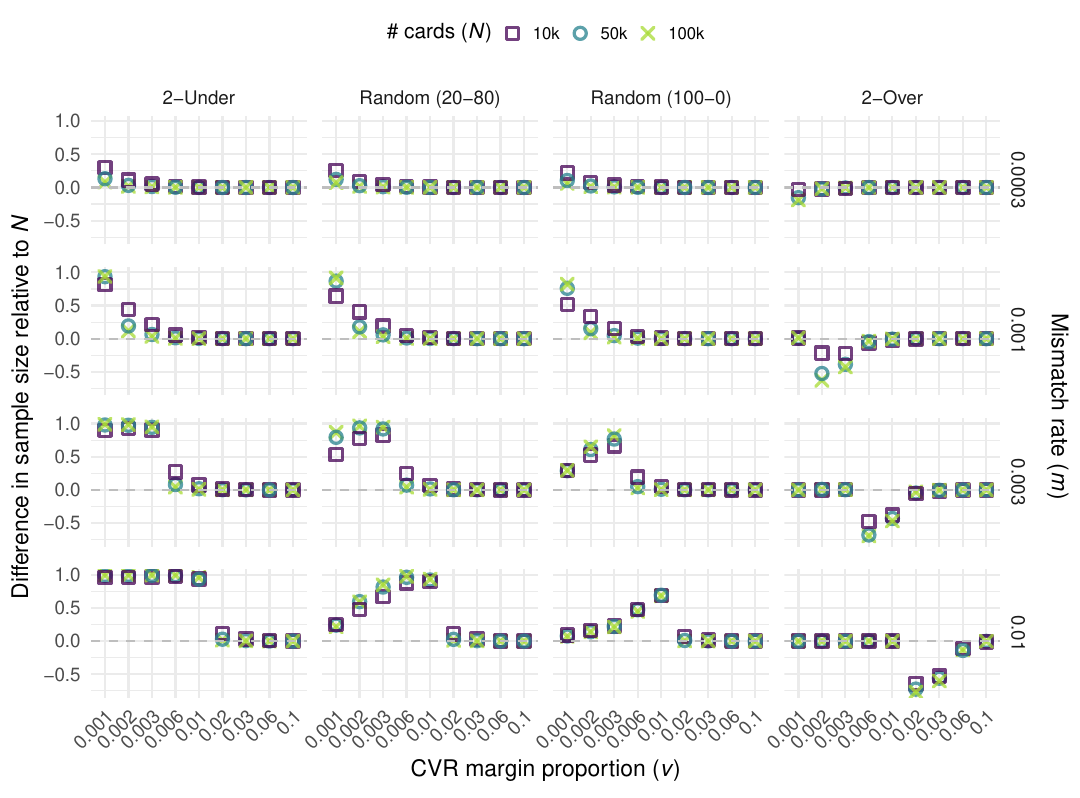}
\caption{\textbf{Cost of \mismatch{} RLAs versus card-level comparison RLAs for
two-candidate plurality contests as a fraction of the number of cards.}
The difference in average sample size (\mismatch{} audit sample size minus card
comparison audit sample size) as a proportion of the total number of CVRs
(y-axis), for various CVR margin proportions $\propmargin$ (x-axis), number
of CVRs $N$ (point type and color; see legend), mismatch rates $m$ (rows,
as labeled on the right), and error models (columns, as labeled at the
top).  Along the y-axis, a value of 0 means the methods gave similar
average sample sizes.  Positive values mean the card-level comparison audit
had smaller average sample size and vice versa for negative values.  Values
near 1 mean that the \mismatch{} RLA required a full hand count when the
card-comparison RLA generally terminated after examining only a small
fraction of the cards.}
\label{fig:plural}
\end{figure}

We examined the cost of \mismatch{} audits compared to card-level comparison
audits of a contest whose outcome is characterized by a single assorter, e.g.,
a two-candidate plurality contest.  We ran experiments similar to those in
\autoref{sec:mismatch-audit-performance}, but also varied the nature of the
discrepancies between $b_i$ and $c_i$, since card-level comparison audits
depend on the types of error as well as their rates.  We considered four error
models:
\begin{description}
\item[2-Under.]
Every error is a 2-vote understatement.
This is gives card-level comparison the greatest advantage.

\item[2-Over.]
Every error is a 2-vote overstatement.
This is the best case for a \mismatch{} audit.

\item[Random (100--0).]
Picking among votes other than $b_i$
(including null votes) uniformly at random.

\item[Random (20--80).]
Same as ``Random (100--0),'' but 80\% of the CVRs have null votes, leading to a
larger probability of having 1-vote over- and understatements.
This scenario assesses the impact of a large number of invalid votes.
For all other scenarios (above), every CVR had a valid vote.
\end{description}
The CVRs were constructed to attain the intended CVR margin proportion and
mismatch rate.  For example, in the ``2-Under'' scenario, the only mismatches
involve CVRs with votes for the losing candidate.

Results are in \autoref{fig:plural}.  When the mismatch rate $m$ is small
compared to the CVR margin proportion $\propmargin$, the relative incremental
cost of \mismatch{} RLAs is low.  As $m$ approaches $\propmargin$, the
incremental cost increases substantially, particularly when not every error is
a two-vote overstatement.  The incremental cost is greatest when $m \approx
\propmargin$ but some CVR errors ``cancel'' (understatements and
overstatements).  Then a \mismatch{} audit will typically lead to a full manual
count, while a card-level comparison audit often certifies the election after
examining only a fraction of the cards.  Note that sometimes the \mismatch{}
audit leads to smaller average sample sizes than the card comparison audit,
when the errors are the best case for the \mismatch{} audit (which are very
unlikely in practice, absent hacking, misconfiguration, or a serious
malfunction). This is possible because we use different assorters and
statistical tests for the two types of audit.

\subsection{IRV Contests}

\begin{figure}[t]
\centering
\includegraphics[width=\linewidth]{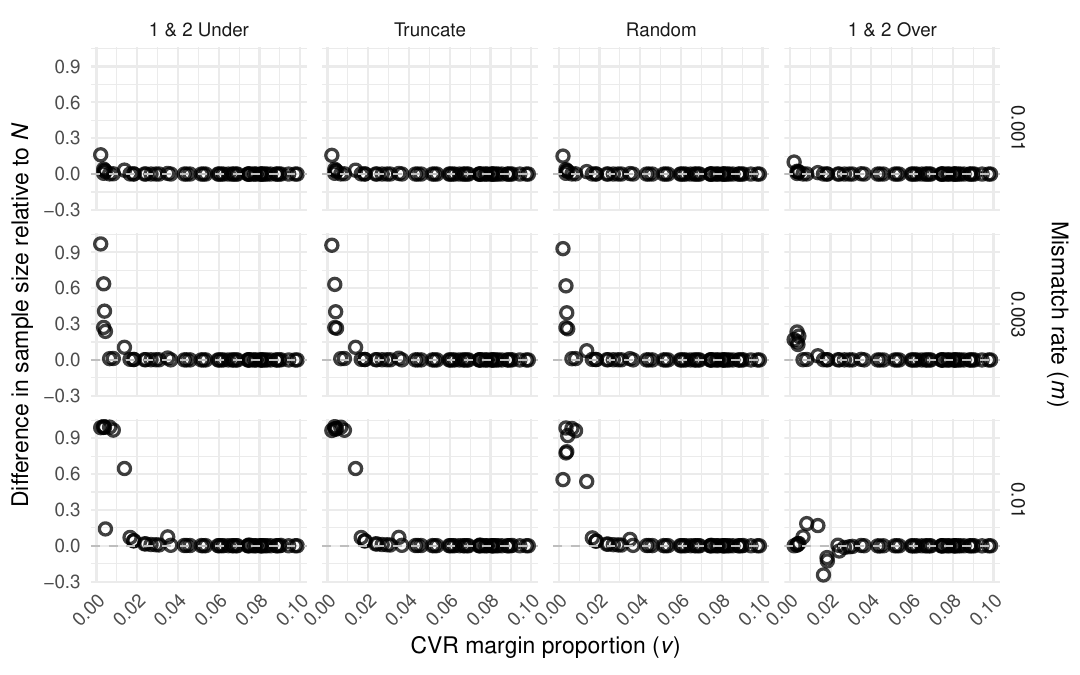}
\caption{\textbf{Relative cost of \mismatch{} RLAs versus RAIRE for IRV
contests.}
Similar to \autoref{fig:plural}, but for IRV rather than two-candidate
plurality.}
\label{fig:irv}
\end{figure}

There are two methods for RLAs for IRV elections: RAIRE~\citep{blom2019raire}
and AWAIRE~\citep{ek2023awaire}.  We compare the sample sizes of \mismatch{}
and RAIRE RLAs.  CVR margins of IRV elections were found using the method of
\citet{ecai2016}.  The simulated audits used the same election data as
\citet{blom2019raire}: 93~New South Wales Legislative Assembly contests and
14~contests in the U.S.\footnote{%
\url{https://github.com/michelleblom/margin-irv/} (visited 26~Jan 2025).}

Results are in \autoref{fig:irv} for four kinds of error:
\begin{description}
\item[Under.]
Every error is a 2-vote (if possible) or a 1-vote (otherwise) understatement of
the assorter with the smallest margin.

\item[Over.]
Every error is a 2-vote (if possible) or a 1-vote (otherwise) overstatement of
the assorter with the smallest margin.

\item[Truncate.]
Truncate the preferences after a position selected uniformly at random between
0 and the number of preferences minus 1.
If the original was a non-vote, randomly rank one candidate first.

\item[Random.]
Replace a vote by randomly choosing ballot length (sampled from the CVRs),
randomly choosing included candidates on the ballot (uniformly), and randomly
permuting these candidates (uniformly), ensuring that the introduced ballot
differs from the CVR.
\end{description}
Performance for IRV elections was similar to performance for two-candidate
plurality elections: for mismatch rates much smaller than the CVR margin
proportion there was essentially no extra cost for a \mismatch{} audit.  When
errors cannot change the election result, a \mismatch{} RLA may require a full
hand count while RAIRE concludes the reported outcome is correct after a
relatively small sample.  When errors are only overstatements of the assorter
with the smallest margin, the most favorable case for \mismatch{} audits,
mismatch-based RLAs
are still generally less efficient than RAIRE.

\subsection{STV Contests}

\begin{table}
\centering
\caption{\Mismatch{} RLA sample sizes (final 3 columns) in simulated audits of
five STV contests in Scotland from 2022.
The number of cards in each contest is $N$.
The \emph{best found lower bound} of the CVR margin is $\cvrmargin^-$ (and
$\propmargin^- = \cvrmargin^- / N$ the corresponding proportion).}
\label{tab:stv}
\begin{tabular}{llccccrrr}
\toprule
Council & Ward & $N$ & Cand. & Seats & $\cvrmargin^-$ ($\propmargin^-$) &
\multicolumn{3}{c}{Mismatch rate ($m$)} \\
\cmidrule{7-9}
 & & \phantom{00,000} & & & & None & 0.0003 & 0.003 \\
\midrule
Glasgow  & Greater Pollok   & 8,869 & 11 & 4 & 161 (1.82\%) & 168 & 178 & 349 \\
Glasgow  & East Centre      & 6,957 & 11 & 4 & 182 (2.62\%) & 115 & 120 & 191 \\
Glasgow  & Drumchapel/A     & 7,226 & 10 & 4 & 323 (4.47\%) &  67 &  68 &  91 \\
Aberdeen & Torry Ferryhill  & 4,997 & 10 & 4 & 254 (5.08\%) &  59 &  60 &  79 \\
Aberdeen & Lower Deeside    & 6,886 & 7  & 3 & 436 (6.33\%) &  47 &  48 &  58 \\
\bottomrule
\end{tabular}
\end{table}

While there is no RLA method for RLAs for STV contests with more than 2
seats~\citep{voting22}, there are methods for computing lower bounds for the
margin of STV
elections~\citep{informsstv} that can handle many more seats.  We simulated
\mismatch{} audits for several STV contests, with three choices for the
mismatch rate.  Our results are in \autoref{tab:stv}.

We believe this is the first example of an efficient RLA for an STV election
with more than 2~seats.  We used the current best approach for computing lower
bounds on CVR margins for STV~\citep{blom2025stvmargin}.

CVR margins for STV are very hard to calculate and the best available lower
bounds sometimes have considerable slack; indeed, the best known lower bound is
0 in some cases.  For the STV contests we considered here, the lower bounds are
a substantial fraction of the number of cards cast, so \mismatch{} audit sample
sizes are small.


\section{Discussion and Conclusion}
\label{sec:conc}

Although they are typically less efficient than more specialized card-level
comparison RLAs, \mismatch{} RLAs have some advantages:
\begin{itemize}
\item They only need a lower bound on the CVR margin for the social choice
    function, which might be possible to compute for some social choice
    functions for which a more nuanced RLA method does not yet exist.
\item It is easier to explain how a \mismatch{} RLA works to a general audience
    than it is to explain, say, RAIRE or AWAIRE.
\end{itemize}
However, when the discrepancy rate is an appreciable fraction of the CVR
margin, \mismatch{} RLAs typically require examining many more ballots,
including requiring a full hand count when a more specialized method would
not.  \Mismatch{} RLAs require a computing (a lower bound on) the CVR margin,
which depends on the social choice function.

Differences in auditing strategies and differences in parameter settings in
ALPHA affect audit sample sizes.  COBRA, which we used for the card-level
comparison audits, does not adapt to the data, so if $m$ differs from the value
assumed by COBRA ($10^{-5}$ in our tests), it could have a disadvantage
compared to the truncated shrinkage estimator used for the \mismatch{} audit,
which ``learns'' the error rate from the sample.  That could explain the small
number of situations where \mismatch{} RLAs require smaller samples than
card-level comparison RLAs: they correspond to using an assumed rate of 2-vote
overstatements in COBRA that is orders of magnitude smaller than the actual
rate of two-vote overstatements.  Furthermore, the estimators for $\eta_j$ that
we used might not be optimally tuned in our applications; further refinements
might improve performance.

\Mismatch{} RLAs can use card-style information to audit a collection of
contests simultaneously and efficiently \cite{glazerEtal21}.  The efficiency of
\mismatch{} RLAs might also be improved by distinguishing among broad classes
of mismatches, for instance, mismatches where either the CVR or the ballot card
has a null vote.  For some social choice functions, it might take, say, $k$
mismatches involving null votes to have the effect a single general mismatch
could have.  For instance, in a plurality contest, a mismatch involving a null
vote can affect the reported margin by at most one vote, rather than two.

\Mismatch{} RLAs require that the CVRs be available to compare to the sampled
ballot cards.  Publishing CVRs may have privacy implications.
\Mismatch{} RLAs are compatible with SOBA \citep{benalohEtal11} and
VAULT \citep{benaloh2019vault}, which use cryptographic commitments to avoid
revealing the full plaintext of CVRs other than those in the RLA sample.


\bibliographystyle{splncsnat}
\bibliography{references}


\end{document}